\documentclass[journal]{IEEEtran}

\usepackage{enumerate}
\usepackage{graphicx}
\usepackage{epsf}
\usepackage{subfigure}
\usepackage{amsmath}
\usepackage{amssymb}
\usepackage{array}
\usepackage{setspace}
\usepackage[amsmath,thmmarks]{ntheorem}
\usepackage[ruled,lined]{algorithm2e}
\usepackage{algorithmic}
\usepackage{color}
\usepackage{amsfonts}
\usepackage{dsfont}
\usepackage{xfrac}
\usepackage{breqn}
\usepackage{bbm}
\usepackage{cite}

\graphicspath{{Fig/},{fig/}}

\theoremseparator{.}

\begin{document}


\title{\LARGE 6D Movable Metasurface (6DMM) in Downlink NOMA Transmissions}


\author{
Li-Hsiang Shen,~\IEEEmembership{Member,~IEEE}}

\maketitle

\begin{abstract}
This letter proposes a novel six-dimensional movable metasurface (6DMM)-assisted downlink non-orthogonal multiple access (NOMA) system, in which a conventional base station (BS) equipped with fixed antennas serves multiple users with the assistance of a reconfigurable intelligent surface (RIS) with six-dimensional spatial configurability. In contrast to traditional RIS with static surface, the proposed 6DMM architecture allows each element to dynamically adjust its position and orient the whole metasurface in yaw-pitch-roll axes, enabling both in spatial and electromagnetic controls. We formulate a sum-rate maximization problem that jointly optimizes the BS NOMA-based beamforming, phase-shifts, element positions, and rotation angles of metasurface under constraints of NOMA power levels, unit-modulus of phase-shifts, power budget, inter-element separation and boundaries of element position/orientation. Due to non-convexity and high-dimensionality, we employ a probabilistic cross-entropy optimization (CEO) scheme to iteratively refine the solution distribution based on maximizing likelihood and elite solution sampling. Simulation results show that the proposed CEO-based 6DMM-NOMA architecture achieves substantial rate performance gains compared to 6DMM sub-structures, conventional static RIS, and other multiple access mechanisms. It also highlights the effectiveness of CEO providing probabilistic optimization for solving high-dimensional scalable metasurface.
\end{abstract}


%

{\let\thefootnote\relax\footnotetext
{Li-Hsiang Shen is with the Department of Communication Engineering, National Central University, Taoyuan 320317, Taiwan. (email: shen@ncu.edu.tw)}}

\section{Introduction}

Reconfigurable intelligent surfaces (RISs) have emerged as a transformative paradigm in the sixth-generation (6G) wireless communication systems, offering a cost-effective and energy-efficient manner to reshape the wireless environment \cite{acm, 3}. By leveraging large arrays of low-cost passive elements capable of adjusting phase-shift control, RISs can dynamically manipulate electromagnetic waves to enhance signal quality, extend coverage, suppress interference, and virtual line-of-sight (LoS) connectivity converting blocked or non-LoS (NLoS) paths. These capabilities have much attracted extensive research in physical layer design, intelligent beamforming, and network-level enhancements to achieve high spectral and energy efficiency \cite{2,dstar}. Despite these advantages, conventional RIS architectures are regarded spatially static, restricting their ability to adapt to rapidly changing channels or user distributions in highly-dynamic environments. 
%

To overcome these limitations, inspiring from the fluid antenna \cite{9, myfas} and movable antenna (MA) structures \cite{MA}, the movable elements (ME) are introduced \cite{lim, movele1} to enable element-wise spatial adjustment across three dimensional (3D) Cartesian axes of x-/y-/z-domain. Different from MAs, ME-enhanced RIS further aligning to proper small-scale channels induces a more challenging task when configuring well-positioned elements in the cascaded channels from the base station (BS), RIS to users \cite{lim1}. Furthermore, with higher degrees of freedom beyond 3D, six-dimensional movable antenna (6DMA) \cite{6dma} is introduced as a revolutionizing technology that exploits the transceiver spatial variation by considering both 3D xyz-positions and 3D rotations, i.e., yaw, pitch and roll, which improves performance in terms of geometric array gains, spatial multiplexing and interference mitigation.

Incorporating the benefits of both 6DMA architectures and RIS, we introduce a six-dimensional movable metasurface (6DMM) as a novel RIS framework that enables element-wise spatial adjustment along the Cartesian axes as well as panel-level rotation in yaw, pitch, and roll angles. Unlike 6DMA, multi-domain reconfigurability of 6DMM provides unprecedented flexibility in dynamically shaping reflected signal waves by movable elements in metasurface. However, enabling such fine-grained reconfiguration at the metasurface level is challenging, as it requires simultaneous geometric and phase alignment between the BS and multiple users under a cascaded channel model, which is different from the 6DMA. From an implementation feasibility perspective, the rotation of the 6DMM structure can be realized by the rotation motor, whereas the spatial positioning can be achieved through either mechanical adjustment \cite{MA} or fluid-style reconfigurable movements \cite{lim}. Note that the tradeoff between the manufacturing cost and performance improvement is left open issues. However, it should be emphasized that the prototype is currently under development, and thus hardware impairment effects are expected to arise from non-ideal mechanical tolerances and circuit implementation constraints \cite{hi,hi2}.

To further enhance spectral efficiency, we consider non-orthogonal multiple access (NOMA) system, superimposing multiuser signals in the power domain at the same frequency \cite{noma}. Moreover, successive interference cancellation (SIC) is employed to decode the user signals according to their channel strengths and power levels. Compared to conventional orthogonal multiple access (OMA) schemes, NOMA offers superior spectral efficiency, user fairness, and connectivity. Integrating NOMA with 6DMM enables a powerful synergy, i.e., NOMA handles interference and user multiplexing in the power domain, whereas 6DMM provides spatial reconfigurability to enhance signal alignment and channel conditions. The main contributions of this letter are elaborated as follows:
\begin{itemize}
    \item We propose a novel 6DMM-assisted downlink NOMA architecture, in which each metasurface element possesses six degrees of freedom, i.e., 3D translation per element and rotational control in yaw-pitch-roll of the entire metasurface. Notably, the BS and users are equipped with conventional fixed antennas.
    
    \item We aim for maximizing sum-rate in the 6DMM-NOMA downlink system, subject to constraints of the transmit power budget at BS, unit-modulus reflection coefficients, inter-element spacing, and bounded space as well as rotational adjustment of 6DMM. To solve the high-dimensional problem, we propose a CEO-based scheme iteratively refining the solution distribution by minimizing the Kullback-Leibler divergence, including the process of Gaussian sampling, fitness evaluation, elite solution selection, parameter updating, and solution smoothing.

\item We have conducted simulations to evaluate the proposed 6DMM-NOMA architecture optimized by CEO scheme. Results demonstrate not only the convergence behavior of CEO parameters but also substantial sum-rate gains compared to reduced 6DMM sub-structures, conventional static RIS, and existing multiple access baselines of OMA and spatial division multiple access (SDMA).

\end{itemize}

\begin{figure}[!t]
\centering
\includegraphics[width=3.3in]{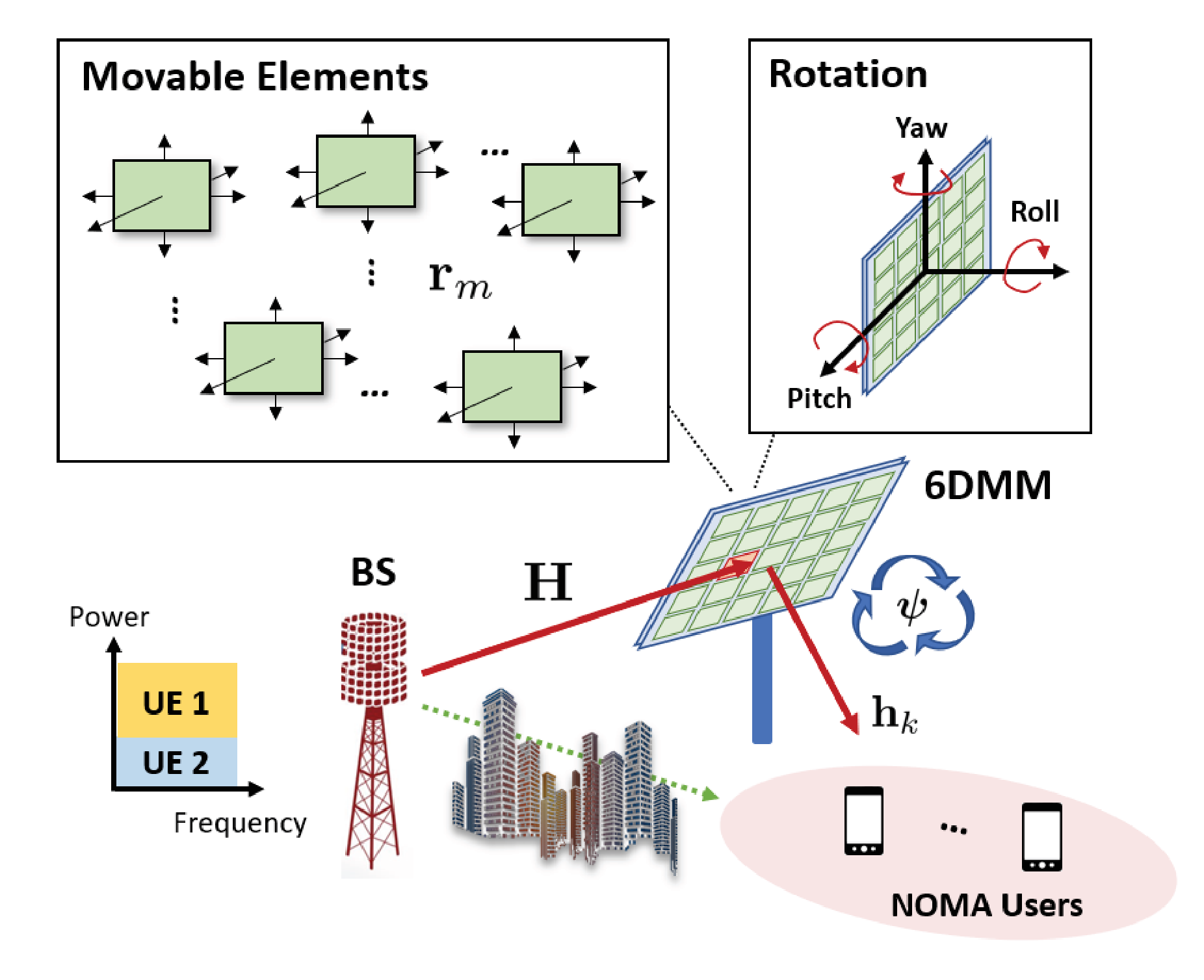}
\caption{The proposed architecture of 6DMM-aided downlink NOMA.} \label{architecture}
\end{figure}

\section{System Model and Problem Formulation}

As shown in Fig. \ref{architecture}, we consider a 6DMM-empowered downlink NOMA in conjunction with a single BS serving $K$ single-antenna users, with its set of $\mathcal{K}=\{1,...,K\}$. The BS is equipped with $N$ transmit antennas, with its set of $\mathcal{N}=\{1,...,N\}$. The 6DMM consists of $M$ movable reflecting elements with its set of $\mathcal{M}=\{1,...,M\}$. Each element position is denoted as $\mathbf{r}_m =[r_{x,m}, r_{y,m}, r_{z,m}]^T$ and its phase-shift is $\theta_m$ where $|\theta_m| = 1$. Notation $\boldsymbol{\theta} = [\theta_1,\dots,\theta_M]^T$ indicates the phase-shift vector of 6DMM, where $T$ is the transpose operation. We consider the Rician fading channel between the BS and 6DMM as 
\begin{align}
\mathbf{H} = \sqrt{ h_0  d_1^{-\alpha}}
\left( \sqrt{\frac{\kappa}{\kappa + 1}} \mathbf{H}_{\text{LoS}} + \sqrt{\frac{1}{\kappa + 1}} \mathbf{H}_{\text{NLoS}} \right),
\end{align}
where $h_0$ is the pathloss at the reference distance of 1 meter, $d_1$ is the distance between the BS and 6DMM, and $\alpha$ is the pathloss exponent. Notation of $\kappa$ is the Rician factor adjusting the portion of $\mathbf{H}_{\text{LoS}}$ and $\mathbf{H}_{\text{NLoS}}$. Then the channel between the 6DMM and user $k$ is defined as $\mathbf{h}_k = \sqrt{ h_0 d_{2,k}^{-\alpha} }
\left( \sqrt{\frac{\kappa}{\kappa + 1}} \mathbf{h}_{k,\text{LoS}} + \sqrt{\frac{1}{\kappa + 1}} \mathbf{h}_{k,\text{NLoS}} \right)$ where $d_{2,k}$ is defined as the distance between 6DMM and user $k$. While $\mathbf{h}_{k,\text{LoS}}$/$\mathbf{h}_{k,\text{NLoS}}$ means LoS/NLoS components of 6DMM to user $k$, respectively. Combined rotation matrix of 6DMM is
$\mathbf{R}(\boldsymbol{\psi}) = 
	\mathbf{R}_z({\psi}_{\text{y}})
	\mathbf{R}_y({\psi}_{\text{p}})
	\mathbf{R}_x({\psi}_{\text{r}})$,
where Euler-angle-based rotation matrices for yaw-pitch-roll in z-y-x sequence is respectively given by $\mathbf{R}_z({\psi}_{\text{y}})$, $\mathbf{R}_y({\psi}_{\text{p}})$ and $\mathbf{R}_x({\psi}_{\text{r}})$ in \eqref{rotate} as shown at top of next page, where ${\psi}_{a}, \forall a \in\{\text{y}, \text{p}, \text{r}\}$ is the angle of each dimension.
\begin{figure*}
\begin{align} \label{rotate}
	\mathbf{R}_z({\psi}_{\text{y}})=
\begin{bmatrix}
\cos {\psi}_{\text{y}} & -\sin {\psi}_{\text{y}} & 0      \\
\sin {\psi}_{\text{y}} & \cos {\psi}_{\text{y}} & 0 \\
0      & 0 & 1
\end{bmatrix}, \
	\mathbf{R}_y({\psi}_{\text{p}})=
\begin{bmatrix}
\cos {\psi}_{\text{p}} & 0 & \sin {\psi}_{\text{p}} \\
0 & 1 & 0 \\
-\sin {\psi}_{\text{p}}  & 0 & \cos {\psi}_{\text{p}}
\end{bmatrix}	, \
	\mathbf{R}_x({\psi}_{\text{r}})=
\begin{bmatrix}
1 & 0 & 0 \\
0 & \cos {\psi}_{\text{r}} & -\sin {\psi}_{\text{r}} \\
0  & \sin {\psi}_{\text{r}} & \cos {\psi}_{\text{r}}
\end{bmatrix}.	
\end{align}
\hrulefill
\end{figure*}
The LoS components contain the array response vectors of BS and 6DMM respectively given by
\begin{align} \label{arrayresponse}
	[\mathbf{a}_{\rm BS}]_n &= \frac{1}{\sqrt{N}} e^{j \frac{2\pi}{\lambda} \mathbf{k}_{\rm BS}^{T} \mathbf{p}_m }, \\
	[\mathbf{a}_{{\rm MM},s}]_m & = \frac{1}{\sqrt{M}} e^{j \frac{2\pi}{\lambda} \mathbf{k}_{{\rm MM},s}^{T} \left[ \mathbf{R} ( \boldsymbol{\psi}) \left( \mathbf{r}_m - \mathbf{c} \right) + \mathbf{c} \right]},
\end{align}
where $s\in \{ r, k \}$ is the index for the angle-of-arrival (AoA) and angle-of-departure (AoD) to user $k$ for 6DMM, and $\lambda$ indicates the wavelength of the operating frequency. Note that $\mathbf{p}_m = (n-1) \cdot d_{\rm BS}$ where $d_{\rm BS}$ is the inter-antenna spacing. The wave vectors of the BS and the 6DMM are denoted by $\mathbf{k}_{\rm BS}$ and $\mathbf{k}_{{\rm MM},s}$, respectively, and follow the general expression $\mathbf{k} = [\sin \vartheta \cos \phi, \sin \vartheta \sin \phi,	 \cos \vartheta]^{\mathrm{T}}$, where $\vartheta$ and $\phi$ denote the azimuth and elevation angles, respectively. Also, $\mathbf{c}$ is the location of whole metasurface. Note that the positional adjustment controls the path-length and phase accumulation, while the rotational adjustment modifies the wavefront orientation and incidence angles, both enabling dynamic channel alternation in a directionally selective manner. Then we can model the LoS parts of BS-6DMM and 6DMM-user respectively as 
\begin{align}
\mathbf{H}_{\text{LoS}} = \mathbf{a}_{{\rm MM},r} \mathbf{a}_{\rm BS}^H, \quad
\mathbf{h}_{k,\text{LoS}} &= \mathbf{a}_{{\rm MM},k},
\end{align}
where $H$ is Hermitian operation. Moreover, the spatial correlation between movable elements in NLoS components should be considered, modeled as the Bessel function of the zero-order as 
\begin{align}
	[\mathbf{C}]_{i,j} = J_0\left( 2\pi d_{i,j} / \lambda \right),
\end{align}
where $d_{i,j} = \lVert \mathbf{r}_i - \mathbf{r}_j \rVert$ is the distance between the element $i$ and $j$ and $\forall i\neq j$. The spatial correlation is further written as $\mathbf{C} = \mathbf{U} \boldsymbol{\Lambda} \mathbf{U}^H$, where $\mathbf{U}$ is unitary matrix with eigenvectors and $ \boldsymbol{\Lambda}$ is the diagonal eigenvalue matrix. Accordingly, the NLoS parts of BS-6DMM and of 6DMM-user can be expressed as $\mathbf{H}_{\text{NLoS}} = \mathbf{C}^{1/2} \bar{\mathbf{H}}$ and 
$\mathbf{h}_{k,\text{NLoS}} = \mathbf{C}^{1/2} \bar{\mathbf{h}}_k$, respectively, where $\mathbf{C}^{1/2} = \boldsymbol{\Lambda}^{1/2} \mathbf{U}^{1/2}$. Note that $\bar{\mathbf{H}}$ and $\bar{\mathbf{h}}_k$ are complex Gaussian distributions both with zero mean and unit variance. Since the relative element locations $d_{i,j}$ remain unchanged before and after the rotation of the whole metasurface, orientation does not influence the correlation. Then the combined channel between the 6DMM and user $k$ is expressed as $\mathbf{g}_k \triangleq \mathbf{h}_{k}^H \boldsymbol{\Theta} \mathbf{H}$, where $\boldsymbol{\Theta} = \mathrm{diag}(\boldsymbol{\theta})$. The received signal of user $k$ is obtained as 
\begin{align}
y_k = \mathbf{g}_k \mathbf{w}_k x_k + \sum_{j \ne k, j\in\mathcal{K}} \mathbf{g}_k \mathbf{w}_j x_j + n_k,
\end{align}
where $\mathbf{w}_k \in \mathbb{C}^{N\times 1}$ is the beamforming vector for user $k$, $x_k \sim \mathcal{CN}(0,1)$ is the transmitted symbol, and $n_k \sim \mathcal{CN}(0,\sigma^2)$ is Gaussian noise of user $k$. Before SIC decoding, we sort out the user index in an ascending order by comparing the channel gain as $|\mathbf{g}_1|^2 \geq |\mathbf{g}_2 |^2 \geq \ldots \geq |\mathbf{g}_K |^2$. During the SIC process in NOMA \cite{mfris}, the stronger user $k$ can decode the signal of the weaker user $k'$ based on the criterion of
\begin{align} \label{con0}
	\frac{| \mathbf{g}_k \mathbf{w}_{k'}|^2}{\sum_{j \ne k',j \in \mathcal{K}} |\mathbf{g}_k \mathbf{w}_j|^2 + \sigma^2} 
	\geq 
	\frac{| \mathbf{g}_k \mathbf{w}_k|^2}{\sum_{j \ne k,j \in \mathcal{K}} |\mathbf{g}_k \mathbf{w}_j|^2 + \sigma^2}.
\end{align}
Considering the perfect SIC cancellation, the signal-to-interference-plus-noise ratio (SINR) of user $k$ is given by
\begin{align} \label{sinr}
	\gamma_k = \frac{| \mathbf{g}_k \mathbf{w}_k|^2}{\sum_{j < k, j \in \mathcal{K}} |\mathbf{g}_k \mathbf{w}_j|^2 + \sigma^2}.
\end{align}
Then the achievable rate for user $k$ is obtained as $R_k = \log_2(1 + \gamma_k)$. We aim for maximizing total downlink rate by optimizing BS beamforming vector $\mathbf{w}_k$, 6DMM phase-shifts $\boldsymbol{\Theta}$, rotation $\boldsymbol{\psi}$ and positions $\mathbf{r}_m$, which is formulated as
\begingroup
\allowdisplaybreaks
\begin{subequations} \label{total_problem}
\begin{align}
& \max_{\substack{\mathbf{w}_k, \boldsymbol{\Theta}, \mathbf{r}_m, \boldsymbol{\psi} }} \quad \sum_{k\in \mathcal{K}} \log_2 \left( 1 + \gamma_k \right) \\
& \quad \text{s.t.}  \quad \eqref{con0}, \quad \sum_{k\in \mathcal{K}}  \|\mathbf{w}_k\|^2 \le P_{\text{th}}, \label{con1} \\
	& \qquad\quad |\theta_m| = 1, \quad \forall m \in \mathcal{M}, \label{con2} \\
	& \qquad\quad  \|\mathbf{r}_m - \mathbf{r}_{m'}\| \ge d_{\text{th}}, \quad \forall m\in \mathcal{M}, m \neq m', \label{con3} \\
	& \qquad\quad\mathbf{r}_m \in \mathcal{B}_r, \quad \forall m \in\mathcal{M}, \label{con4} \\
	&  \qquad\quad \boldsymbol{\psi} \in \mathcal{B}_{\boldsymbol{\psi}}. \label{con5}
\end{align}
\end{subequations}
\endgroup
In \eqref{con1}, maximum power is constrained by $P_{\text{th}}$. \eqref{con2} limits the phase-shifts of 6DMM. Constraint \eqref{con3} confines the inter-element spacing of 6DMM by $d_{\text{th}}$. \eqref{con4} limits the element positions within the metasurface $\mathcal{B}_r$, where $\mathcal{B}_r = \{ \mathbf{r}_{\text{min}} \preceq \mathbf{r}_m \preceq \mathbf{r}_{\text{max}} |\forall m\in\mathcal{M}\}$ associated with its boundary between $\mathbf{r}_{\text{min}}$ and $\mathbf{r}_{\text{max}}$. Constraint \eqref{con5} indicates the available range of yaw-pitch-roll as $\mathcal{B}_{\boldsymbol{\psi}} = \{ {\psi}_{a,\text{min}} \preceq {\psi}_a \preceq {\psi}_{a,\text{max}} | \forall a\in\{\text{y}, \text{p}, \text{r} \} \}$, associated with its boundary between ${\psi}_{a,\text{min}}$ and ${\psi}_{a,\text{max}}$. Due to the non-convex nature in problem \eqref{total_problem}, conventional optimization techniques might struggle to search the global optimum. To address this challenge, we design a CEO-based approach in a stochastic probability manner, which is suitable for solving high-dimensional and non-convex problem. 

\section{Proposed CEO-based Scheme}

The CEO scheme is an iterative stochastic search technique designed to efficiently explore high-dimensional or complex solution spaces \cite{ceo}. It operates by maintaining and updating a parametrized probability distribution over the solution space, aiming to progressively concentrate sampling in regions that yield higher objective values. At each iteration $t$, the CEO scheme proceeds through the following steps:

\subsubsection{{Sampling}} 
We define the individual solution as $\mathbf{x}_i = [\mathbf{x}_{\mathfrak{R}}, \mathbf{x}_{\mathfrak{I}}, \mathbf{x}_{\theta}, \mathbf{x}_{r}, \mathbf{x}_{\psi} ] \in \mathbb{R}^{ D}$, corresponding to the parameters $\{\mathbf{w}_k, \boldsymbol{\Theta}, \mathbf{r}_m, {\psi}_a \}$ and its dimension $D=2NK + M + 3M + 3$. Note that the beamforming is $\mathbf{x}_{\mathfrak{A}} = {\rm Flat}(\mathfrak{A}[\mathbf{W}])$, where $\mathfrak{A}\in \{ \mathcal{\mathfrak{R}, \mathfrak{I}} \}$ indicates the real/imaginary parts and ${\rm Flat}(\cdot)$ reshapes the matrix $\mathbf{W} = [ \mathbf{w}_1, ..., \mathbf{w}_K]$ into a 1-dim vector. The others notations are defined as $\mathbf{x}_{\theta} = [\varphi_1, ..., \varphi_M]$, $\mathbf{x}_{r} = [r_{x,1}, r_{y,1}, r_{z,1},... ,r_{x,M}, r_{y,M}, r_{z,M}]$, and $\mathbf{x}_{\psi} = [ \psi_{\text{y}}, \psi_{\text{p}}, \psi_{\text{r}}]$. Then a population of $I$ candidate solutions $\{ \mathbf{x}_1, ..., \mathbf{x}_I \}$ is drawn from a parametric probability distribution $p(\mathbf{x}; \boldsymbol{\Xi}^{(t)})$, where $\mathcal{I}=\{1,...,I\}$ is an index set and $\boldsymbol{\Xi}^{(t)}=\{\boldsymbol{\mu}^{(t)}, \boldsymbol{\varpi}^{(t)} \}$ denotes the distribution parameters of mean and variance in the considered Gaussian case, with its joint probability density function (PDF) composed of $D$ independent Gaussian distributions 
\begin{align}
 p(\mathbf{x} \in \mathbf{x}_i; \boldsymbol{\mu}, \boldsymbol{\varpi}) = \prod_{d=1}^{D} \frac{1}{\sqrt{2\pi} \varpi_{d}} e^{- \frac{(x_{d} - \mu_{d})^2}{2 \varpi_{d}^2}},
\end{align}    
where $ \boldsymbol{\mu} = [\mu_1, ..., \mu_D ]$ is and $\boldsymbol{\varpi} = \{\varpi_1, ..., \varpi_D\}$ indicate the means and variance, respectively. To elaborate further, the sampled parameters $\mathbf{x}_i$ will be processed according to their boundary constraints: The beamforming $\mathbf{w}_k$ is normalized to sustain its original distribution but is satisfied with its power constraint, i.e., $\mathbf{w}_k = \frac{\mathbf{w}_k}{\sum_{k\in \mathcal{K}}  \|\mathbf{w}_k\|^2} \cdot P_{\text{th}}$. The phase-shift of 6DMM is defined as $\theta_m = e^{-j \varphi_m}$ with $\varphi_m\in ( 0, 2\pi]$. The boundary of each element position is $\mathbf{r}_m = \min \left( \max (\mathbf{r}_m, \mathbf{r}_{\text{min}}), \mathbf{r}_{\text{max}} \right)$, whilst the yaw-pitch-roll is updated as ${\psi}_a = \min \left( \max ({\psi}_a, {\psi}_{a,\text{min}}), {\psi}_{a,\text{max}} \right)$.

\subsubsection{{Evaluation}} 
The objective function is evaluated for each sampled solution $\mathbf{x}_i$, which is designed as
\begin{align} \label{fit}
	f(\mathbf{x}_i) &= \sum_{k\in \mathcal{K}} R_k \!-\! \varrho  \sum_{m\in\mathcal{M}} \sum_{m'\in\mathcal{M} \backslash m} \left( d_{\text{th}} \!-\! \|\mathbf{r}_m \!-\! \mathbf{r}_{m'}\| \right),
\end{align}
where $\varrho \leftarrow \varrho / \sqrt{t}$ is the decaying penalty factor. 


\subsubsection{{Elite Selection}} 
A subset of samples of top fitness values evaluated based on \eqref{fit} is selected, termed as the \textit{elite} set $\mathcal{E}^{(t)}$. Then elite set comprises the top $\rho \in (0,1]$ quantile of samples with the highest objective values, with its size defined as the cardinality of the set as $|\mathcal{E}^{(t)}| = \lceil \rho \cdot I \rceil$. We sort out the solution set $\mathbf{x}_i$ with the top fitness values in an descending order, i.e., $f(\mathbf{x}_1) \geq f(\mathbf{x}_2) \geq ... \geq f(\mathbf{x}_i) \geq f(\mathbf{x}_j) \geq ... \geq f(\mathbf{x}_{|\mathcal{E}^{(t)}|})$. Then the elite set is given by 
\begin{align} \label{elite}
    \mathcal{E}^{(t)} = \left\{ \mathbf{x}_i \;\middle|\; f(\mathbf{x}_i) \geq f(\mathbf{x}_j), \forall i,j \in\mathcal{I}, \forall i\neq j  \right\}.
\end{align}

\subsubsection{{Parameter Update}} 
 The distribution parameters $\boldsymbol{\Xi}$ are updated by minimizing the Kullback-Leibler (KL) divergence, i.e., cross-entropy between the current distribution and the empirical distribution of the elite set, which is equivalent to solving $\boldsymbol{\Xi}^{(t+1)} = \arg\max_{\boldsymbol{\Xi}} \sum_{\mathbf{x}_i \in \mathcal{E}^{(t)}} \log p(\mathbf{x}_i; \boldsymbol{\Xi})$. This update shifts the sampling distribution toward regions with higher-quality solutions. Then the updated sampled mean and variance can be respectively acquired as
\begin{align}
\mu_d^{(t+1)}  \!=\! \frac{\sum\limits_{i\in\mathcal{I}} [\mathbf{x}_{i}]_d }{|\mathcal{E}^{(t)}|} , \ \ \varpi_d^{(t+1)} \!=\! \sqrt{\frac{ \sum\limits_{i\in\mathcal{I}} \left( [\mathbf{x}_{i}]_d \!-\! \mu_d^{(t+1)} \right)^2  }{|\mathcal{E}^{(t)}|}}.
\end{align}
Note that the derived sampled mean and variance are optimal based on the maximization of the average log-likelihood. The proof can be found in \cite{LLproof}.

\subsubsection{{Smoothing}} 
To improve stability and prevent premature convergence, parameter smoothing is designed as 
\begin{align} \label{smooth}
    \boldsymbol{\Xi}^{(t+1)} \leftarrow \varsigma \boldsymbol{\Xi}^{(t+1)} + (1 - \varsigma) \boldsymbol{\Xi}^{(t)},
\end{align}
where $\varsigma \in (0, 1]$ is the smoothing factor.

\subsubsection{{Termination}} 
Conduct procedures above and update the temporary optimum $f^{*}(\mathbf{x}^{*}) = f(\mathbf{x}_1^{(t)})$ with $\mathbf{x}^{*} = \mathbf{x}_1^{(t)}$ at $t$-th iteration until satisfaction of $| f(\mathbf{x}_1^{(t)}) - f^{*}(\mathbf{x}^{*}) | \leq f_{\text{th}}$, where $f_{\text{th}}$ is the predefined threshold. The CEO iteratively adapts its sampling distribution to promising solution regions, effectively balancing exploration and exploitation. Note that sampling solution distribution of CEO progressively concentrates around the optimum as the iterations evolve. 
The solution generation of CEO is in a complexity order of $\mathcal{O}(I\cdot (NK+M))$, whereas elite set is with an order of $\mathcal{O}(I^2)$. The parameter  updates of mean and variance are both with an order of $\mathcal{O}(I\cdot (NK+M))$, whilst the smoothing is in a polynomial order $\mathcal{O}(1)$. As a result, the overall complexity order is $\mathcal{O}(T_{\text{th}} \cdot (I^2 + INK + IM)$, where $T_{\text{th}}$ is the iteration bound.

\section{Simulation Results}

\begin{figure*}[!t]
	\centering
	\subfigure[]{\includegraphics[width=3 in]{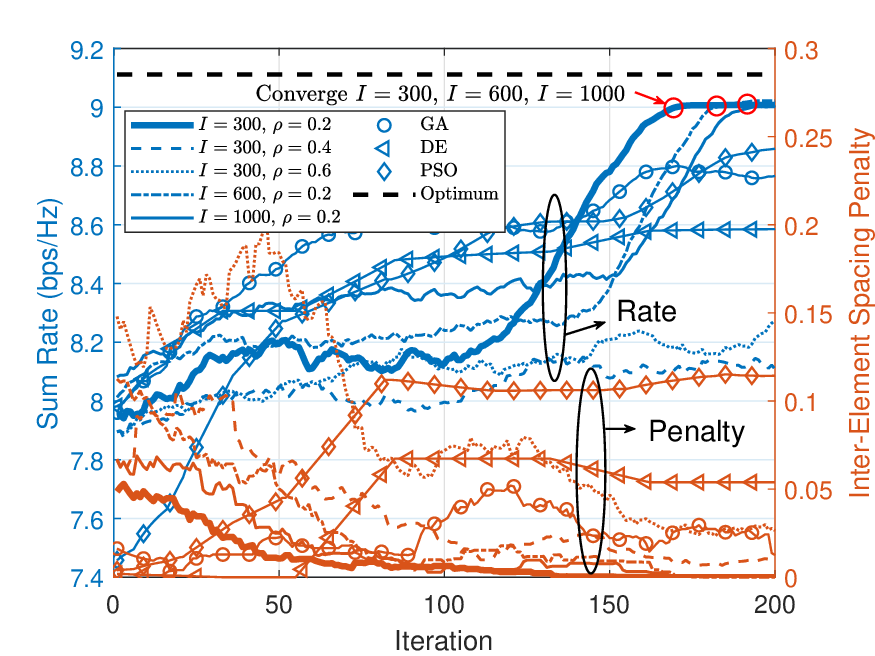} \label{fig1}}
	\subfigure[]{\includegraphics[width=3 in]{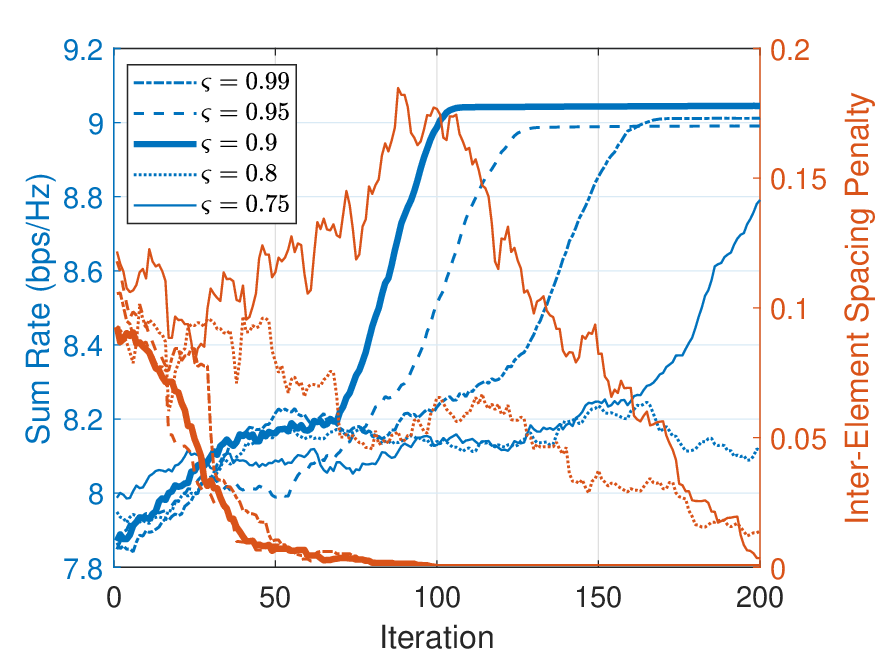} \label{fig1-1}}
	\subfigure[]{\includegraphics[width=3 in]{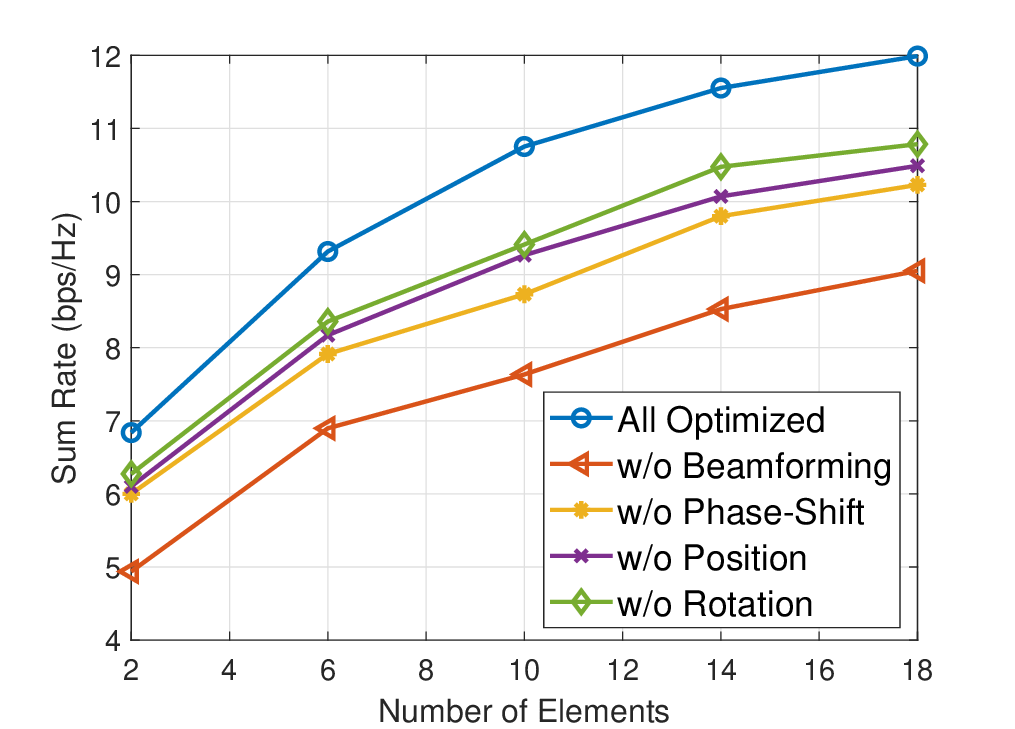} \label{fig2}}
	\subfigure[]{\includegraphics[width=3 in]{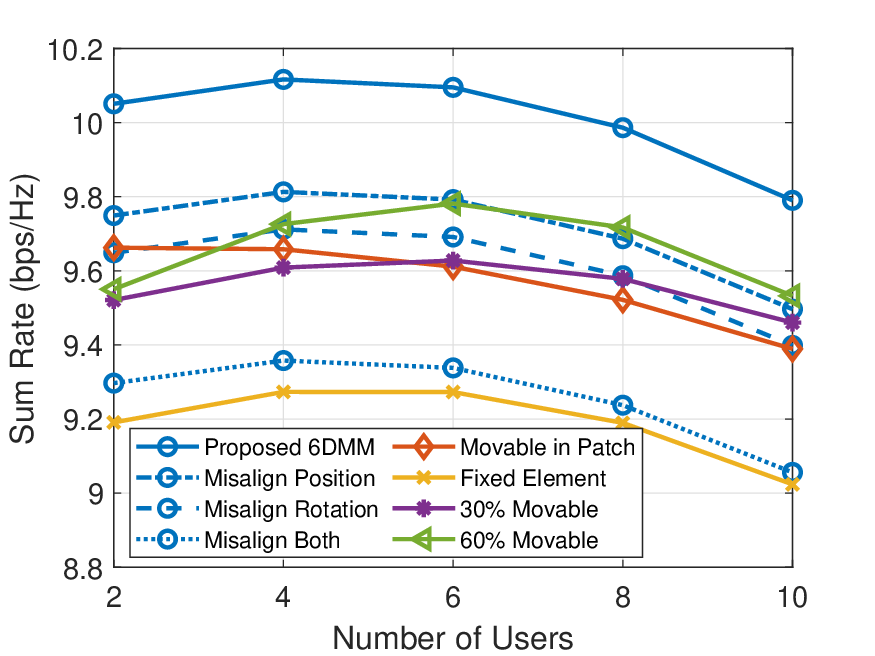} \label{fig3}}
	\subfigure[]{\includegraphics[width=3 in]{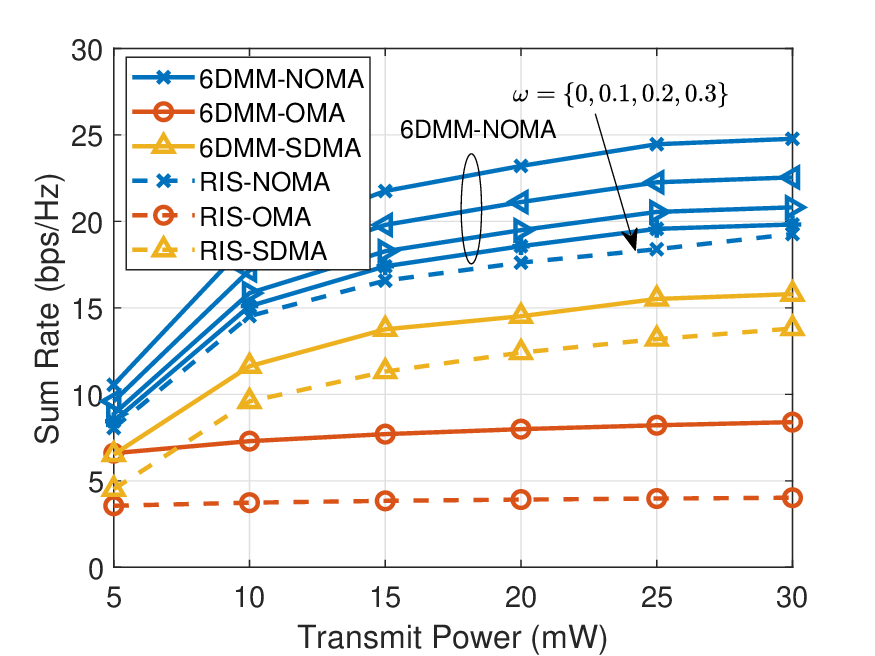} \label{fig4}}
	\caption{
	(a) Convergence with different $I$ and $\rho$. 
	(b) Convergence with different $\varsigma$. 
	(c) Rate with different numbers of elements. 
	(d) Rate with various numbers of users. 
	(e) Rate versus different maximum power.}
\end{figure*}

The proposed CEO solution for 6DMM-NOMA is evaluated through simulations, where pertinent parameter settings can be referred from \cite{movele1, lim1, LLproof}. The BS and 6DMM are located at $(0,0,10)$ and $(50, 20, 10)$ m, respectively. Users are uniformly and randomly distributed within a radius of $10$ m centered at the location $(100, 0, 2)$ m. The number of antennas/elements/users are set to $N=16$, $M=16$, and $K=4$, respectively. The position boundary \cite{movele1} of each movable element is set to
$\mathbf{r}_{\text{min}} = [0,0,0]$ and
$\mathbf{r}_{\text{max}} = [1,1,1]$, i.e., volume of $1$ m$^3$, whereas its rotation is constrained by $\psi_{\text{r},\text{min}} = -\pi$, $\psi_{\text{r},\text{max}} = \pi$, $\psi_{\text{p},\text{min}} = -\pi/3$, $\psi_{\text{p},\text{max}} = \pi/3$, $\psi_{\text{y},\text{min}} = -\pi/3$, and $\psi_{\text{y},\text{max}} = \pi/3$. The other related parameters are set as follows: $h_0 = -20$ dB, $\alpha = 2.2$, $\kappa=3$ , $\lambda = 0.1$ m, $\sigma^2 = -95$ dBm, $P_{\text{th}}=5$ mW, $d_{\text{th}} = 0.05$ m \cite{lim1}, $I=300$, $\varrho=0.2$, $\varsigma=0.9$ and $T_{\text{out}}=200$ \cite{LLproof}.


Fig.~\ref{fig1} illustrates the rate convergence behavior of the proposed CEO-based solution for 6DMM-NOMA with $I\in\{300,600,1000\}$ and $\rho\in\{0.2, 0.4, 0.6\}$. It can be observed that the case with $I = 300$ samples and $\rho=0.2$ converges faster than other cases, as it has fewer samples for elite selection and parameter updates. However, although the case with $I = 600$ exhibits slower convergence, it results in an asymptotic rate to that of $I=300$ owing to the possible inclusion of suboptimal solutions. Notably, the case with $I = 1000$ achieves the same rate as the case with $I = 300$, but suffers from the slowest convergence and incurs nearly three times higher computational complexity per iteration. Therefore, $I=300$ is selected when evaluating the impact of $\rho\in\{0.2, 0.4, 0.6\}$. It is observed that a larger elite selection ratio $\rho$ tends to include more suboptimal samples, such as those yielding low rates or violating the separation constraint, which significantly degrades the overall system performance. As a result, the cases with $\rho \in \{ 0.4, 0.6\}$ fail to converge under large penalty factors due to the dominance of infeasible or unqualified solutions in the elite set. The performance of the CEO scheme closely approaches the global optimum obtained via brute-force search, exhibiting less than a $2\%$ rate gap. Moreover, we compare CEO to the genetic algorithm (GA) \cite{GA} and differential evolution (DE), both evolving a population through elite selection, crossover, and mutation operations. We also benchmark particle swarm optimization (PSO) \cite{pso} updating solutions using local and global best positions with velocity dynamics. We observe that all GA/DE/PSO benchmarks lack strict convergence guarantees and are susceptible to premature convergence due to stochastic operators potentially trapped in local optima. Particularly, GA and DE share similar evolutionary dynamics and thus often exhibit correlated convergence behaviors, whilst PSO tends to converge more slowly due to swarm contraction. Moreover, Fig.~\ref{fig1-1} depicts the impact of selecting the smoothing factor $\varsigma$. We can observe that smaller $\varsigma$ has quicker convergence but with worse rate performance and penalty guarantee. While, too large $\varsigma$ results in a moderate rate and slower convergence. Accordingly, selecting suitable $\varsigma$ becomes essential.

Fig.~\ref{fig2} illustrates the rate performance versus the number of elements under different optimization configurations. As expected, the sum rate increases monotonically with the number of elements due to enhanced array gain and spatial diversity. As the number of elements increases, the performance gaps between these cases become more evident, emphasizing the importance of joint optimization of 6DMM configurations. The curve with the \textit{all optimized} case jointly optimizes all parameters by CEO, performing the highest rate among all the other cases. Comparatively, the performance degrades when any individual parameters are not optimized. Note that all configurations are optimized using the proposed CEO framework, while the "w/o" case serves as a baseline where randomization is applied instead of optimization. Specifically, the case without beamforming exhibits the most significant performance loss, highlighting the critical role of transmit beamforming alignment to 6DMM in achieving high rate. The cases without phase-shift and without position also incur noticeable performance drops, suggesting that both passive reflection control and spatial configuration of elements considerably impact the system throughput. Moreover, the case without rotation exhibits moderate rate degradation, indicating that metasurface rotation is less dominant than phase-shift or position adaptation due to limited degree of freedom.

Fig.~\ref{fig3} presents the sum rate performance with different numbers of users under various 6DMM structures. We observe an initial increase in the rate due to dynamic reconfigurability, followed by a decline as the system reaches a limit in its degrees of freedom. The proposed 6DMM achieves the highest sum rate across all cases, benefited from its full reconfigurability. In contrast, the conventional RIS with \textit{fixed elements} shows the lowest rate due to inadaptability to user distribution and channel dynamics, having a rate degradation of around $10\%$ compared to 6DMM. The structure \textit{movable in patch} indicates that each element is only permitted to move inside its corresponding patch without inter-patch migration. While the movable element within patch-array structure introduces adaptability and can yield moderate performance gains by fine-tuning the element positions locally, the improvement remains limited due to the constrained movement radius. The cases of $30\%, 60\%$ movable elements offer intermediate rate between fixed and fully optimized cases due to their semi-configurability. Moreover, more movable elements outperform the patch-array strategy when $K=4$ to $7$ users, confirming the advantages of partially adaptive metasurfaces. Additionally, we evaluate the impacts of the mechanical misalignment in positional/rotational adjustment with respective errors of $0.005$ m and $\pi/90$ rad. It is inferred that the either imperfect position or rotation adjustment will lead to declined rates, reaching almost the similar rate level to that of the partially-movable structure. Intriguingly, the rate with both imperfections will result in the asymptotic rate to that of the conventional fixed element based metasurfaces.

Fig.~\ref{fig4} illustrates the sum rate performance versus transmit power under different multiple access schemes of NOMA, OMA and SDMA as well as metasurfaces. As transmit power increases, all schemes exhibit an improved rate due to enhanced signal strength. The proposed 6DMM-NOMA consistently achieves the highest rate across all cases, leveraging both spatial adaptability and the power-domain multiplexing of NOMA. Comparably, 6DMM-SDMA performs a moderate rate without the benefit of interference cancellation. Meanwhile, 6DMM-OMA shows the lowest sum rate as it underutilizes available resources. For the RIS-based benchmarks, a rate performance degradation is clearly observed compared to 6DMM structure, owning to the limited reconfigurability of movable metasurface and elements. RIS-NOMA outperforms RIS-SDMA and RIS-OMA, yet falls below 6DMM counterparts due to the limited degree of freedom of conventional fixed RIS. Moreover, we additionally consider the interference $\omega \cdot \sum_{j > k, j \in \mathcal{K}} |\mathbf{g}_k \mathbf{w}_j|^2$ in SINR to quantify the imperfect SIC, where $\omega$ means the imperfection level. We can observe that the rate is declined with higher $\omega$ values in NOMA case due to the existence of residual strong user interferences. Note that the benefit of 6DMM-NOMA under imperfect SIC approaches that of RIS-NOMA when $\omega=0.3$ which indicates the importance of SIC mechanism.

\section{Conclusion}

In this letter, we have proposed a novel 6DMM-assisted downlink NOMA transmission framework, which enables full electromagnetic control of transmit beamforming and phase-shift as well as spatial reconfigurability through element positions and metasurface orientations. By jointly optimizing the NOMA-based beamforming vectors and the 6D configurations of the metasurface, the system potentially achieves enhanced adaptability to user distribution and channel conditions. To efficiently solve the resulting high-dimensional and non-convex rate maximization problem, we adopt the CEO algorithm relying on the maximizing log-likelihood, which probabilistically explores the solution space and iteratively updates distribution parameters based on elite samples. Simulation results validate the effectiveness of the proposed 6DMM-NOMA with CEO scheme. The 6DMM accomplishes the highest rate among the cases of movable elements in patches, partially-movable, and fixed elements. 6DMM-NOMA also shows consistent rate gains over conventional RIS-based systems and different multiple access strategies of SDMA and OMA. The results also highlight the significant potential of CEO solving the problem combining high-dimensional metasurface with NOMA for achieving high rates.

\bibliographystyle{IEEEtran}
\bibliography{IEEEabrv}
\end{document}